\documentclass[letterpaper,10pt,twocolumn,oneside,final,conference]{IEEEtran}

\usepackage{amsfonts}
\usepackage{amsmath}
\usepackage{amssymb}
\usepackage{graphicx}
\usepackage{cite}

\ifCLASSINFOpdf
\else
\fi
\begin{document}

\title{Electron-Acoustic Solitons in an Electron-Beam Plasma System with 
kappa-distributed Electrons }

\author{\IEEEauthorblockN{A.~Danehkar}
\IEEEauthorblockA{Department of Physics and Astronomy,\\ Macquarie
University,\\ Sydney, NSW 2109, Australia\\
Email: ashkbiz.danehkar@students.mq.edu.au}
\and
\IEEEauthorblockN{I.~Kourakis}
\IEEEauthorblockA{Department of Physics and Astronomy,\\ Queen's
University Belfast, Belfast,\\ BT7 1NN, United Kingdom}
\and
\IEEEauthorblockN{M.~A.~Hellberg}
\IEEEauthorblockA{School of Physics, University of \\KwaZulu-Natal, Private Bag X54001,
\\ Durban 4000, South Africa}}

\maketitle
\thispagestyle{plain}
\pagestyle{plain}

\begin{abstract}
We investigate the existence conditions and propagation properties of electron-acoustic solitary waves in a plasma consisting of an electron beam fluid, a cold electron fluid, and a hot suprathermal electron component modeled by a $\kappa$-distribution function. The Sagdeev pseudopotential method was used to investigate the occurrence of stationary-profile solitary waves. We have determined how the soliton characteristics depend on the electron beam parameters. It is found that the existence domain for solitons becomes narrower with an increase in the suprathermality of hot electrons, increasing the beam speed, and decreasing the beam-to-cold electron population ratio.
\end{abstract}

\markboth{IEEE 41st International Conference on Plasma Sciences (ICOPS),
May~2014, DOI: 10.1109/PLASMA.2014.7012747}{Danehkar \MakeLowercase{\textit{et al.}}: Electron-Beam Plasma
System with $\kappa$-distributed Electrons}


\IEEEpeerreviewmaketitle

\section{Introduction}

\IEEEPARstart{I}{nteraction} of a stream of high-energy electrons with the
background plasma plays an important role in the astrophysical phenomena
such as solar bow shock \cite{Cane2002,Klein2005,Reiner2008} and Earth's
foreshock emission \cite{Cairns1988,Cairns1994}. Electron beams can emerge
directly as a fast stream of electrons propagating through the background
plasma, or indirectly from electrons accelerated by slow propagating
hydrodynamic shocks. It is not yet fully understood how electrostatic
solitary waves are produced at the bow shock.

Interestingly, a population of energetic suprathermal electrons was also found
to exist in those environments, which has a suprathermal tail on the velocity
distribution function \cite{Vasyliunas1968}. Energetic electrons are often
modeled by a $\kappa$-distribution function having high-energy tails of the
suprathermal (non-Maxwellian) forms \cite{Vasyliunas1968}. The suprathermality
is identified by the spectral index $\kappa$, which describes how it deviates
from a Maxwellian. Low values of $\kappa$ are associated with a significant
suprathermality, whereas Maxwellian distribution is recovered in the limit
$\kappa\rightarrow\infty$. The common form of the $\kappa$-velocity
distribution function is given by \cite{Summers1991,Baluku2008,Hellberg2009}:%
\begin{equation}
f_{\kappa}(v)=n_{0}(\pi\kappa\theta^{2})^{-3/2}\frac{\Gamma(\kappa+1)}%
{\Gamma(\kappa-\frac{1}{2})}\left(  1+\frac{v^{2}}{\kappa\theta^{2}}\right)
^{-\kappa-1}. \label{eq1_60}%
\end{equation}
where $n_{0}$ is the equilibrium number density of the electron, $v$ the
velocity variable, and $\theta=v_{th,e}\left[  (\kappa-\tfrac{3}{2}%
)/\kappa\right]  ^{1/2}$ the most probable speed related to the usual thermal
velocity $v_{th,e}=(2k_{B}T_{e}/m_{e})^{1/2}$. Here, $k_{B}$ is the Boltzmann
constant, $m_{e}$ the electron mass, and $T_{e}$ the temperature of an
equivalent Maxwellian having the same energy content. The term involving the
Gamma function $\Gamma$ arises from the normalization of $f_{\kappa}(v)$,
viz., $\int f_{\kappa}(v)d^{3}v=n_{0}$. The spectral index describes the
suprathermality, with $\kappa>3/2$ for reality.

In the previous work \cite{Danehkar2011,Danehkar2011b,Saini2011}, we have studied the properties of
negative electrostatic potential solitary structures exist in a plasma with
excess suprathermal electrons. In the present work, we aim to study the
existence conditions and propagation properties of electron-acoustic solitary
waves in a plasma consisting of an electron beam fluid, a cold electron fluid,
and hot suprathermal electrons modeled by a $\kappa$-distribution function.

\section{Theoretical model}

We consider a plasma consisting of four components, namely a cold inertial
drifting electron-fluid (the beam), a cold inertial background electron-fluid,
an inertialess hot suprathermal electron component modeled by a $\kappa
$-distribution, and uniformly distributed stationary ions.

The cold electron behavior is governed by the following normalized
one-dimensional equations,
\begin{align}
&  \frac{\partial n}{\partial t}+\frac{\partial(nu)}{\partial x}%
=0,\label{eq_11}\\
&  \frac{\partial u}{\partial t}+u\frac{\partial u}{\partial x}=\frac
{\partial\phi}{\partial x}, \label{eq_12}%
\end{align}
and for the electron beam,%
\begin{align}
&  \frac{\partial n_{b}}{\partial t}+\frac{\partial(n_{b}u_{b})}{\partial
x}=0,\label{eq_14}\\
&  \frac{\partial u_{b}}{\partial t}+u_{b}\frac{\partial u_{b}}{\partial
x}=\frac{\partial\phi}{\partial x}, \label{eq_15}%
\end{align}
Here, $n$ and $n_{b}$ denote the fluid density variables of the cold electrons
and the beam electrons normalized with respect to the equilibrium number
density of cold electron-fluid $n_{c,0}$ and electron beam $n_{b,0}$,
respectively. The velocities $u$ and $u_{b}$, and the equilibrium beam speed
$U_{0}=u_{b,0}/c_{th}$ are scaled by the hot electron thermal speed
$c_{th}=\left(  k_{B}T_{h}/m_{e}\right)  ^{1/2}$, and the wave potential
$\phi$ by $k_{B}T_{h}/e$. Time and space are scaled by the plasma period
$\omega_{pc}^{-1}=\left(  n_{c,0}e^{2}/\varepsilon_{0}m_{e}\right)  ^{-1/2}%
$\ and the characteristic length $\lambda_{0}=\left(  \varepsilon_{0}%
k_{B}T_{h}/n_{c,0}e^{2}\right)  ^{1/2}$, respectively, where $\varepsilon_{0}$
is the permittivity constant and $T_{h}$ is the temperature of the hot electrons.

The following normalized $\kappa$-distribution is adopted for the number
density of the hot electrons \cite{Summers1991,Baluku2008,Hellberg2009}:
\begin{equation}
n_{h}=\alpha\left(  1-\frac{\phi}{(\kappa-\tfrac{3}{2})}\right)
^{-\kappa+1/2}. \label{eq_17}%
\end{equation}
where $\alpha=n_{h,0}/n_{c,0}$ is the hot-to-cold electron charge density
ratio, $n_{h,0}$ the equilibrium number density of hot electrons.

The ions are assumed to be immobile in a uniform state, so $n_{i}=n_{i,0}=$
const, where $n_{i,0}$ is the undisturbed ion density. At equilibrium, the
plasma is quasi-neutral, so $Zn_{i,0}=n_{c,0}+n_{b,0}+n_{h,0}$. We also define
the beam-to-cold electron charge density ratio $\beta=n_{b,0}/n_{c,0} $, so
$Z{n_{i,0}}/{n_{c,0}}=1+\alpha+\beta$.

All four components are coupled via the Poisson's equation as follows
\begin{equation}
\frac{\partial^{2}\phi}{\partial x^{2}}=-\left(  1+\alpha+\beta\right)
+n+\beta n_{b}+n_{h}. \label{eq_18}%
\end{equation}

\section{Linear waves}

\begin{figure}
[ptb]
\begin{center}
\includegraphics[
width=3.0in
]%
{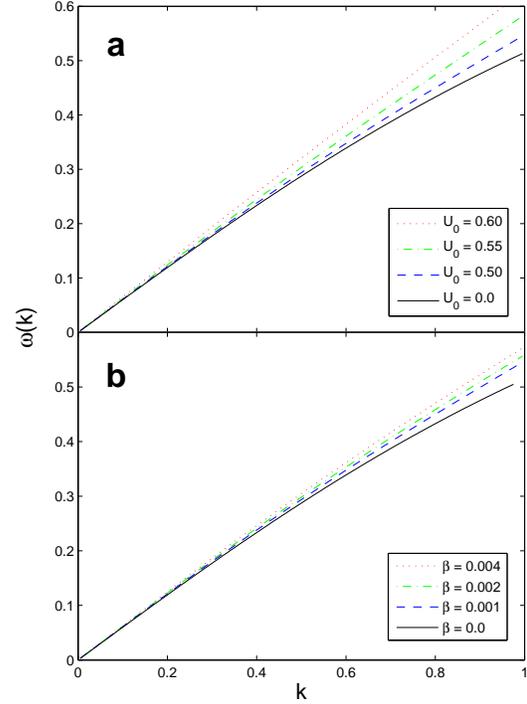}%
\caption{Dispersion curve for harmonic (linear) electron-acoustic waves. Upper
panel (a): The variation of the dispersion curve for different values of
\ $U_{0}$. Curves from bottom to top: $U_{0}=0$ \ (solid), $0.5$\ (dashed),
$0.55$\ (dot-dashed), and $0.6$\ (dotted curve). Bottom panel (b): Variation
of the dispersion curve for different values of \ $\beta$. Curves from bottom
to top: $\beta=0.0$\ (solid), $0.001$\ (dashed), $0.002$\ (dot-dashed curve),
and $0.004$\ (dotted curve). Here, (a) $\beta=0.001$ (b) $U_{0}=0.5$, and
(a-b) $\alpha=1$ and $\kappa=3$. }%
\label{fig1}%
\end{center}
\end{figure}

As a first step, we consider linearized forms of Eqs. (\ref{eq_11}%
)-(\ref{eq_15}) to study small-amplitude harmonic waves of frequency $\omega$
and wavenumber $k$. We assume that $S=\{n,u,n_{b},u_{b},\phi\}$ describes the
system's state at a given position $x$ and instant $t$. A small deviation from
the equilibrium state $S^{(0)}=\{1,0,1,U_{0\text{ }},0\}$ by taking
$S=S^{(0)}+S_{1}^{(1)}e^{i(kx-\omega t)}$ leads to the derivatives of the
first order amplitudes $\partial S_{1}^{(1)}/\partial t=-i\omega S_{1}^{(1)}%
$and $\partial S_{1}^{(1)}/\partial x=ikS_{1}^{(1)}$. Using these derivatives,
we obtain the following equations:%
\begin{align}
n_{1}^{(1)} &  =\dfrac{k}{\omega}u_{1}^{(1)},\text{ \ \ }u_{1}^{(1)}%
=-\dfrac{k}{\omega}\phi_{1}^{(1)},\label{eq_19}\\
&  n_{b1}^{(1)}=\dfrac{k}{\omega-U_{0\text{ }}k}u_{b1}^{(1)},\label{eq_20}\\
u_{b1}^{(1)} &  =-\dfrac{k}{\omega}\left(  \phi_{1}^{(1)}-U_{0\text{ }}%
u_{b1}^{(1)}\right)  ,\label{eq_21}%
\end{align}
Substituting Eqs. (\ref{eq_19})--(\ref{eq_21}) to the Poisson's equation
(\ref{eq_18}) and make use of the expansion keeping up to first order provides
the following linear dispersion relation%
\begin{equation}
1+\frac{k_{D,\kappa}^{2}}{k^{2}}-\frac{1}{\omega^{2}}-\frac{\beta}{\left(
\omega-kU_{0\text{ }}\right)  ^{2}}=0.\label{eq_25}%
\end{equation}
The appearance of a normalized $\kappa$-dependent screening factor
$k_{D,\kappa}$ in the denominator, is defined by
\begin{equation}
k_{D,\kappa}\equiv\dfrac{1}{\lambda_{D,\kappa}}\equiv\left[  \dfrac
{\alpha(\kappa-\frac{1}{2})}{\kappa-\tfrac{3}{2}}\right]  ^{1/2}%
\,.\label{eq_24}%
\end{equation}
Figure \ref{fig1} shows the effect of varying the values of the electron beam
velocity $U_{0}$ and the beam-to-cold electron charge density ratio $\beta
$\ on the dispersion curve. As seen, the phase speed ($\omega/k$) increases
weakly with an increase in the electron beam parameters $U_{0}$ and $\beta$.
An increase in the number density of suprathermal hot electrons or the
suprathermality (decreasing $\kappa$) also decreases the phase speed, in
agreement with what found in Ref. \cite{Danehkar2011}.%

\section{Nonlinear analysis}

To obtain solitary wave profile solutions, we consider all fluid variables in
a stationary frame traveling at a constant normalized velocity $M$ (to be
referred to as the Mach number), implying the transformation $\xi=x-Mt$. This
replaces the space and time derivatives with $\partial/\partial x=d/d\xi$ and
$\partial/\partial t=-Md/d\xi$, respectively. Now equations (\ref{eq_11}%
)-(\ref{eq_15}) and (\ref{eq_18}) take the form:
\begin{align}
&  -M\dfrac{dn}{d\xi}+\frac{d(nu)}{d\xi}=0,\label{eq_26}\\
&  -M\dfrac{du}{d\xi}+u\dfrac{du}{d\xi}=\dfrac{d\phi}{d\xi}, \label{eq_27}%
\end{align}%
\begin{align}
&  -M\dfrac{dn_{b}}{d\xi}+\frac{d(n_{b}u_{b})}{d\xi}=0,\label{eq_29}\\
&  -M\dfrac{du_{b}}{d\xi}+u_{b}\dfrac{du_{b}}{d\xi}=\dfrac{d\phi}{d\xi},
\label{eq_30}%
\end{align}%
\begin{align}
\dfrac{d^{2}\phi}{d\xi^{2}}  &  =-\left(  1+\alpha+\beta\right)  +n+\beta
n_{b}\nonumber\\
&  +\alpha\left(  1-\frac{\phi}{(\kappa-\tfrac{3}{2})}\right)  ^{-\kappa+1/2},
\label{eq_32}%
\end{align}
The equilibrium state is assumed to be reached at both infinities
($\xi\rightarrow\pm\infty$), so integrating Eqs. (\ref{eq_26})--(\ref{eq_30})
and applying the boundary conditions $n=1$, $u=0$, $n_{b}=1$, $u_{b}%
=U_{0\text{ }}$ and $\phi=0$ at infinities provide%
\begin{align}
&  u=M\left[  1-\left(  \frac{1}{n}\right)  \right]  ,\label{eq_33}\\
&  u={M-}\left(  {M}^{2}{+2\phi}\right)  ^{1/2},\label{eq_34}\\
&  u_{b}=M\left[  1-\frac{1}{n_{b}}\left(  1-\frac{U_{0\text{ }}}{M}\right)
\right]  ,\label{eq_35}\\
&  u_{b}=M-\left(  M^{2}+2\phi-2MU_{0\text{ }}+U_{0}^{2}\right)  ^{1/2},
\label{eq_36}%
\end{align}
Combining Eqs. (\ref{eq_33})--(\ref{eq_36}), one obtains the following
equations for the cold electron density and beam electron density,
respectively%
\begin{equation}
n=\left(  1+\dfrac{2\phi}{M^{2}}\right)  ^{-1/2}, \label{eq_40}%
\end{equation}%
\begin{equation}
n_{b}=\left(  1+\dfrac{2\phi}{(M-U_{0\text{ }})^{2}}\right)  ^{-1/2},
\label{eq_41}%
\end{equation}
Substituting the density expression (\ref{eq_40}) and (\ref{eq_41}) into
Poisson's equation (\ref{eq_32}) and integrating, yields a pseudo-energy
balance equation:
\begin{equation}
\frac{1}{2}\left(  \frac{d\phi}{d\xi}\right)  ^{2}+\Psi(\phi)=0, \label{eq_42}%
\end{equation}
where the Sagdeev pseudopotential $\Psi(\phi)$ is given by%
\begin{align}
\Psi(\phi)  &  =\beta(M-U_{0\text{ }})^{2}\left[  1-\left(  1+\dfrac{2\phi
}{(M-U_{0\text{ }})^{2}}\right)  ^{1/2}\right] \nonumber\\
&  +\left(  1+\alpha+\beta\right)  \phi+M^{2}\left[  1-\left(  1+\dfrac{2\phi
}{M^{2}}\right)  ^{1/2}\right] \nonumber\\
&  +\alpha\left[  1-\left(  1-\frac{\phi}{(\kappa-\tfrac{3}{2})}\right)
^{-\kappa+3/2}\right]  . \label{eq_43}%
\end{align}
In the absence of the beam ($\beta\rightarrow0$), Eq. (\ref{eq_43}) recovers
Eq. (33) given in Ref. \cite{Danehkar2011} in the cold-electron limit
($T_{e}=0$).

\begin{figure}
[ptb]
\begin{center}
\includegraphics[
width=3.0in
]%
{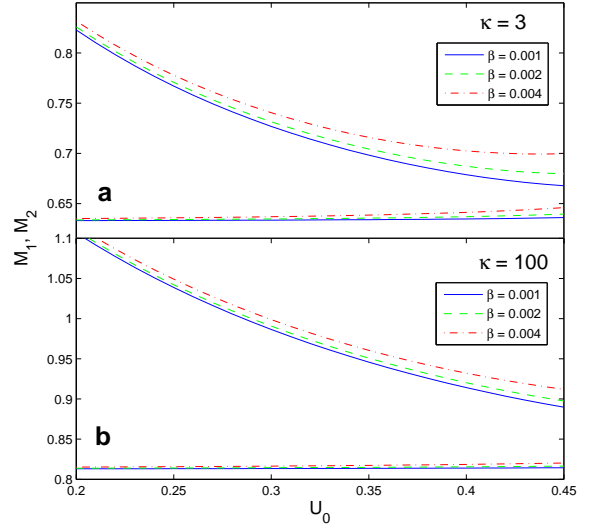}%
\caption{Variation of the lower limit $M_{1}$ (lower curves) and the upper
limit $M_{2}$ (upper curves) with the equilibrium beam speed $U_{0}$ for
different values of the beam-to-cool electron charge density ratio $\beta$.
Curves: $\beta=0.001$\ (solid), $0.002$\ (dashed), and $0.004$\ (dot-dashed).
Here,  (a) $\kappa=3$, (b) $\kappa=100$ and (a-b) $\alpha=1.5$.}%
\label{fig2}%
\end{center}
\end{figure}

\section{Soliton existence domain}

\begin{figure}
[ptb]
\begin{center}
\includegraphics[
width=3.0in
]%
{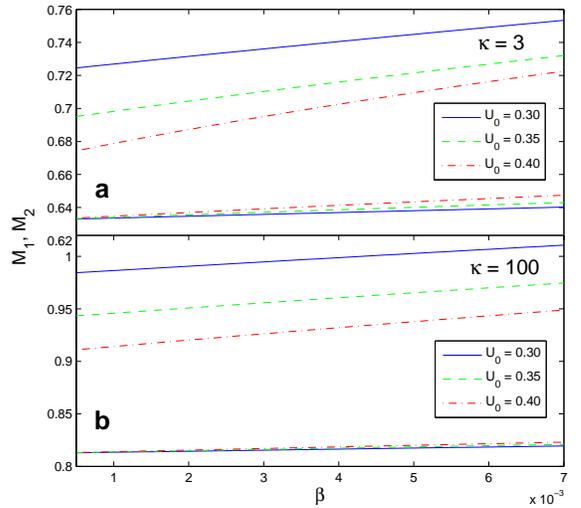}%
\caption{Variation of the lower limit $M_{1}$ (lower curves) and the upper
limit $M_{2}$ (upper curves) with the beam-to-cool electron charge density
ratio $\beta$ for different values of the equilibrium beam speed $U_{0}$.
Curves: $U_{0}=0.3$\ (solid), $0.35$\ (dashed), and $0.4$\ (dot-dashed). Here,
we have taken: (a) $\kappa=3$ and (b) $\kappa=100$. Here, we have taken: (a)
$\kappa=3$, (b) $\kappa=100$ and (a-b) $\alpha=1.5$.}%
\label{fig3}%
\end{center}

For existence of solitons, we require that the origin at $\phi=0$ is a root
and a local maximum of $\Psi$ in Eq. (\ref{eq_43}), i.e., $\Psi(\phi)=0$,
$d\Psi(\phi)/d\phi=0$ and $d^{2}\Psi(\phi)/d\phi^{2}<0$ at $\phi=0$. The first
two constraints are satisfied. We thus impose the condition $F_{1}%
(M)=-d^{2}\Psi(\phi)/d\phi^{2}|_{\phi=0}>0$, and we get%
\begin{equation}
F_{1}(M)=\frac{\alpha(\kappa-\frac{1}{2})}{\kappa-\tfrac{3}{2}}-\frac{1}%
{M^{2}}-\frac{\beta}{(M-U_{0\text{ }})^{2}}>0 \label{eq_44}%
\end{equation}
Eq. (\ref{eq_44}) provides the minimum value for the Mach number $M_{1}$.

\end{figure}
\begin{figure}
[ptb]
\begin{center}
\includegraphics[
width=3.1in
]%
{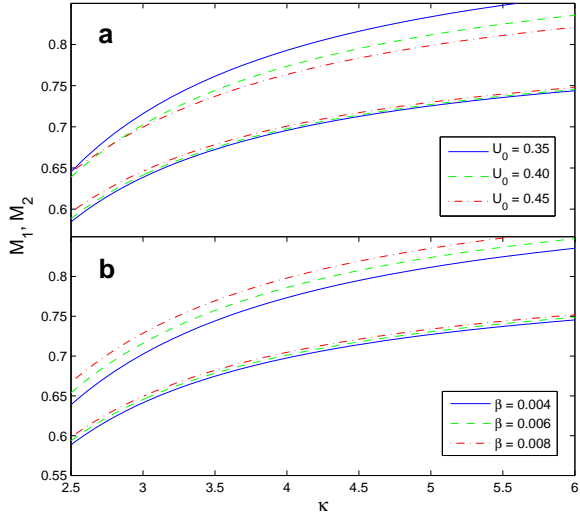}%
\caption{Variation of the lower limit $M_{1}$ (lower curves) and the upper
limit $M_{2}$ (upper curves) with the suprathermality parameter $\kappa$ for
different values of the equilibrium beam speed $U_{0}$ (a) and the
beam-to-cool electron charge density ratio $\beta$ (b). Upper panel (a):
$U_{0}=0.35$\ (solid curve), $0.40$\ (dashed), and $0.45$\ (dot-dashed).
Middle panel (b): $\beta=0.004$\ (solid curve), $0.006$\ (dashed), and
$0.008$\ (dot-dashed). Here, we have taken (a) $\beta=0.004$, (b) $U_{0}=0.4$
and (a-b) $\alpha=1.5$.}%
\label{fig4}%
\end{center}
\end{figure}

An upper limit for M is found through the fact that the cold electron density
becomes complex at $\phi_{\lim(-)}=-\frac{1}{2}{M}^{2}$ for $U_{0}\leqslant0$
and $\phi_{\lim(-)}=-\frac{1}{2}\left(  M-U_{0}\right)  ^{2}$ for $U_{0}>0$,
which yield the following equations for the upper limit in $M$ for
$U_{0}\leqslant0$:%
\begin{align}
F_{2}(M)  &  =M^{2}\left(  1-\tfrac{1}{2}\left(  1+\alpha+\beta\right)
\right) \nonumber\\
&  +\beta(M-U_{0\text{ }})^{2}\left[  1-\left(  1-\dfrac{M^{2}}{(M-U_{0\text{
}})^{2}}\right)  ^{1/2}\right] \nonumber\\
&  +\alpha\left[  1-\left(  1+\dfrac{M^{2}}{2\kappa-3}\right)  ^{-\kappa
+3/2}\right]  \label{eq_45}%
\end{align}
and for $U_{0}>0$:%
\begin{align}
F_{2}(M)  &  =-\tfrac{1}{2}\left(  1+\alpha-\beta\right)  (M-U_{0}%
)^{2}\nonumber\\
&  +M^{2}\left[  1-\left(  1-\dfrac{(M-U_{0})^{2}}{M^{2}}\right)
^{1/2}\right] \nonumber\\
&  +\alpha\left[  1-\left(  1+\dfrac{(M-U_{0})^{2}}{2\kappa-3}\right)
^{-\kappa+3/2}\right]  \label{eq_46}%
\end{align}
Solving equations (\ref{eq_45}) and (\ref{eq_46}) provides the upper limit
$M_{2}$ for acceptable values of the Mach number for solitons to exist. In the
absence of the beam ($\beta\rightarrow0$), Eqs. (\ref{eq_44}) and
(\ref{eq_45}) yield exactly Eqs. (34) and (36) given in Ref.
\cite{Danehkar2011} in the cold-electron limit ($T_{e}=0$).

Figure \ref{fig2} depicts the existence domain of electron-acoustic solitary
waves in two opposite cases: a very low, and a very high value of $\kappa$. We
see that the existence domain in Mach number becomes narrower for strong
suprathermality and higher values of the equilibrium beam speed $U_{0}$. From
two frames (a) and (b) in Fig. \ref{fig2}, it is found that low value of
$\kappa$ imposes that the soliton propagates at lower Mach number range. We
note that lower values of the beam-to-cold electron charge density ratio
($\beta$; see also Fig. \ref{fig3}) shrink the permitted soliton region for
very high $U_{0}$ ($\geqslant0.5$) and strong suprathermality (low $\kappa$).%

As seen in Figs. \ref{fig2} and \ref{fig3}, the existence region becomes
narrower for lower values of $\beta$ and $\kappa$. It is in contrast to
increasing the hot-to-cold electron charge density ratio $\alpha$, which
shrinks down the existence region \cite{Danehkar2011}. As seen, a high value
of the beam speed $U_{0}$ shrinks the permitted region for strong
suprathermality (low $\kappa$).%

\begin{figure}
[ptb]
\begin{center}
\includegraphics[
width=3.1in
]%
{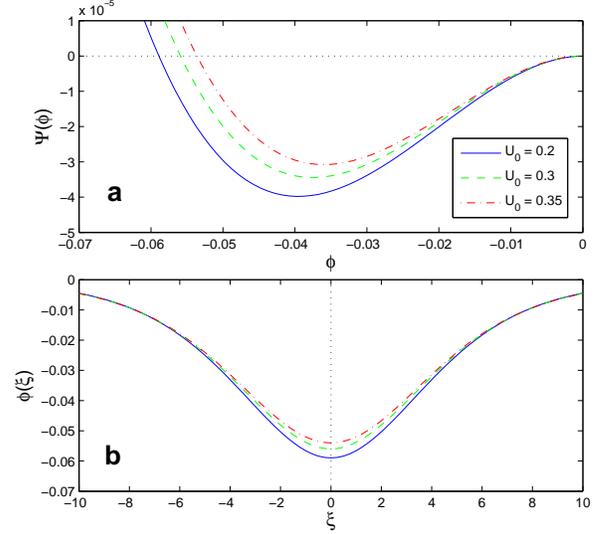}%
\caption{The pseudopotential $\Psi(\phi)$ vs. $\phi$\ (a) and the associated
electric potential pulse $\phi$ vs. $\xi$\ (b) for different values of the
equilibrium beam speed $U_{0}$. From bottom to top: $U_{0}=0.2$\ (solid
curve), $0.3$\ (dashed curve), and $0.35$\ (dot-dashed curve). Here, we have
taken: $\alpha=1$, $\beta=0.008$, $\kappa=4.0$\ and $M=0.9$.}%
\label{fig5}%
\end{center}
\end{figure}

Figure \ref{fig4} shows the effect of a $\kappa$-distribution of hot
electrons. The acoustic limits ($M_{1}$ and $M_{2}$) decreases rapidly as
approaching the limiting value $\kappa\rightarrow3/2$. However, going towards
a Maxwellian distribution ($\kappa\rightarrow\infty$) broadens the permitted
range of the Mach number. The result is similar to the trend in Figs.
\ref{fig2} and \ref{fig3}. It is also similar to what we found in the model
without the beam \cite{Danehkar2011}.%

\section{Soliton characteristics}

\begin{figure}
[ptb]
\begin{center}
\includegraphics[
width=3.1in
]%
{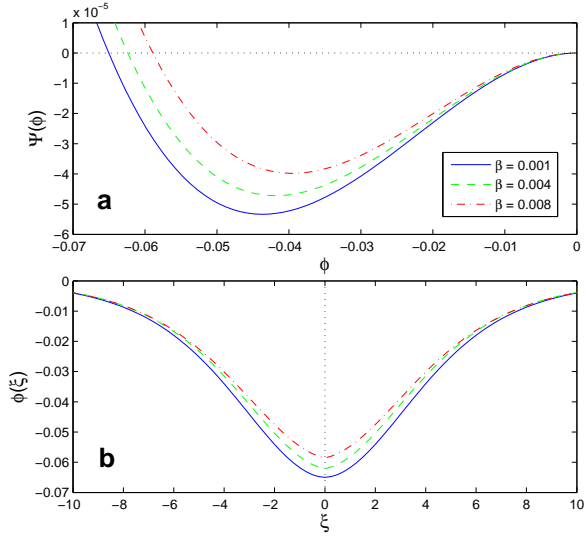}%
\caption{The pseudopotential $\Psi(\phi)$ vs. $\phi$\ (a) and the associated
electric potential pulse $\phi$ vs. $\xi$\ (b) for different values of the
beam-to-cool electron charge density ratio $\beta$. From bottom to top:
$\beta=0.001$\ (solid curve); $0.004$ (dashed curve); $0.008$ (dot-dashed
curve). Here, we have taken $\alpha=1$, $U_{0}=0.2$, $\kappa=4.0$\ and
$M=0.9$.}%
\label{fig6}%
\end{center}
\end{figure}

Figure \ref{fig5}a shows the variation of the pseudopotential $\Psi(\phi)$
with the normalized potential $\phi$, for different values of the beam speed
$U_{0}$ (keeping $\alpha=1$, $\beta=0.008$, $\kappa=4$ and $M=0.9$). The
electrostatic pulse (soliton) solution shown in Fig. 5b is obtained via
numerical integration. As seen, the pulse amplitude $|\phi_{m}|$ decreases
with increasing $U_{0}$.%

Figure \ref{fig6} shows the variation of the pseudopotential $\Psi(\phi)$ for
different values of the beam-to-cold electron charge density ratio $\beta$.
Both the root and the depth of the Sagdeev potential increase with decreasing
$\beta=n_{b,0}/n_{c,0}$. This means that either increasing the cold electron
density or decreasing the electron beam density increase the negative
potential solitary waves.

\section{Conclusion}

In the present study, we have investigated the linear and nonlinear
large-amplitude characteristics of electron-acoustic solitary waves in a
plasma consisting of electron beam, hot $\kappa$-distributed electrons, cold
background electrons and immobile ions.
We derived the linear dispersion
relation of our model, and determined the effects of beam parameters on the
dispersion characteristics, namely the beam-to-cool electron population ratio
$\beta$ and the equilibrium beam speed $U_{0}$. 
We have used the Sagdeev
pseudopotential method to investigate large amplitude localized nonlinear
electrostatic structures (solitary waves), and to determine the region in
parameter space where stationary profile solutions may exist. We have found
only negative potential solitons as apparently the $\kappa$-distribution does
not lead to reverse polarity. The existence domain for solitons was found to
become narrower with an increase in the suprathermality (decreasing $\kappa$),
increasing the beam speed $U_{0}$, decreasing the beam-to-cold electron
population ratio $\beta$. We numerically obtained a series of appropriate
examples of the electrostatic solitons, which also supports the soliton
permitted regions obtained through a root and a local maximum of the
pseudopotential. Our results will improve the understanding of solitary waves
observed in space electron-beam plasmas, which often include energetic
suprathermal electrons.

\section*{Acknowledgment}

Work supported by Macquarie University Research Excellence Scholarship (MQRES) and UK Engineering and Physical Science Research Council (EPSRC grant No. EP/D06337X/1).

\ifCLASSOPTIONcaptionsoff
\newpage \fi



\end{document}